\documentclass[twocolumn,showpacs,preprintnumbers,superscriptaddress,amsmath,amssymb]{revtex4}
\usepackage{algorithm}
\usepackage{algpseudocode}
\usepackage[all]{xy}
\usepackage{amsfonts}
\usepackage{amsmath}
\usepackage{amssymb}
\usepackage{anysize}
\usepackage{bm}
\usepackage{braket}
\usepackage{dcolumn}
\usepackage{epsf}
\usepackage{graphicx}
\usepackage{hyperref}
\usepackage{mathrsfs}
\usepackage[amsmath,thmmarks]{ntheorem}
\usepackage{subfigure}

\newtheorem{theorem}{Theorem}
\newtheorem{example}{Example}
\newtheorem{corollary}{Corollary}
\newtheorem{definition}{Definition}

\newtheorem{proposition}{Proposition}
\usepackage{algpseudocode}
\newcommand{\bthm}{\begin{theorem}}
	\newcommand{\ethm}{\end{theorem}}
\newcommand{\blem}{\begin{lemma}}
	\newcommand{\elem}{\end{lemma}}
\newcommand{\bex}{\begin{example}}
	\newcommand{\eex}{\end{example}}
\newcommand{\bprop}{\begin{proposition}}
	\newcommand{\eprop}{\end{proposition}}
\newcommand{\bplm}{\begin{problem}}
	\newcommand{\eplm}{\end{problem}}
\newcommand{\bmrk}{\begin{remark}}
	\newcommand{\emrk}{\end{remark}}
\newcommand{\bdfn}{\begin{definition}}
	\newcommand{\edfn}{\end{definition}}
\newcommand{\bcor}{\begin{corollary}}
	\newcommand{\ecor}{\end{corollary}}

\newcommand{\beq}{\begin{equation}}
\newcommand{\eeq}{\end{equation}}
\newcommand{\beqm}{\begin{equation*}}
\newcommand{\eeqm}{\end{equation*}}
\newcommand{\beqn}{\begin{eqnarray}}
\newcommand{\eeqn}{\end{eqnarray}}
\newcommand{\beqnm}{\begin{eqnarray*}}
	\newcommand{\eeqnm}{\end{eqnarray*}}
\newcommand{\bea}{\begin{align}}
\newcommand{\eea}{\end{align}}
\newcommand{\beam}{\begin{align*}}
\newcommand{\eeam}{\end{align*}}
\newcommand{\bs}{\begin{subequations}}
	\newcommand{\es}{\end{subequations}}

\newcommand{\bei}{\begin{itemize}}
	\newcommand{\eei}{\end{itemize}}
\newcommand{\bed}{\begin{description}}
	\newcommand{\eed}{\end{description}}
\newcommand{\bee}{\begin{enumerate}}
	\newcommand{\eee}{\end{enumerate}}

\newcommand{\bey}{\begin{array}}
	\newcommand{\eey}{\end{array}}

\newcommand{\beb}{\begin{thebibliography}{99}}
	\newcommand{\eeb}{\end{thebibliography}}

\usepackage{color}

\begin{document}
	\title{Quantum Optimal Control Theory for the Shaping of Flying Qubits}
	\author{Xue Dong}
	\affiliation{Center for Intelligent and Networked Systems, Department of Automation, Tsinghua University, Beijing, 100084, China}

    \author{Xi Cao}
	\affiliation{Center for Intelligent and Networked Systems, Department of Automation, Tsinghua University, Beijing, 100084, China}

    \author{Wen-Long Li}
	\affiliation{Center for Intelligent and Networked Systems, Department of Automation, Tsinghua University, Beijing, 100084, China}

	\author{Guofeng Zhang}\email{guofeng.zhang@polyu.edu.hk}
	\affiliation{Department of Applied Mathematics, The Hong Kong Polytechnic University, Hung Hom, Kowloon, Hong Kong, China}

    \author{Zhihui Peng}\email{zhihui.peng@hunnu.edu.cn}
	\affiliation{Department of Physics and Synergetic Innovation Center for Quantum Effects and Applications, Hunan Normal University, Changsha 410081, China}

	\author{Re-Bing Wu}\email{rbwu@tsinghua.edu.cn}
	\affiliation{Center for Intelligent and Networked Systems, Department of Automation, Tsinghua University, Beijing, 100084, China}
    \email{rbwu@tsinghua.edu.cn}

	\date{\today}
	
	\begin{abstract}
		The control of flying qubits carried by itinerant photons is ubiquitous in quantum networks. Beside their logical states, the shape of flying qubits must also be tailored for high-efficiency information transmission. In this paper, we introduce quantum optimal control theory to the shaping of flying qubits. Building on the flying-qubit control model established in our previous work, we design objective functionals for the generation of shaped flying qubits under practical constraints on the emitters and couplers. Numerical simulations employing gradient-descent algorithms demonstrate that the optimized control can effectively mitigate unwanted level and photon leakage caused by these non-idealities. Notably, while coherent control offers limited shaping capacity with a fixed coupler, it can significantly enhance the shaping performance when combined with a tunable coupler that has restricted tunability. The proposed optimal control framework provides a systematic approach to achieving high-quality control of flying qubits using realistic quantum devices.
	\end{abstract}
	
	\maketitle
\section{Introduction}

The rapid development of quantum information processing technologies has attracted extensive attention from academia and industries~\cite{Ladd2010}. In the near future, quantum information processing units are to be interconnected for secure communication and distributed quantum computing~\cite{Wei2022}. Central to this vision, the efficient control of flying qubits is vital for high-fidelity quantum information transmission over quantum networks~\cite{divincenzo1995quantum,cirac1997quantum}.

Flying qubits can be physically carried by itinerant photonic fields~\cite{houck2007generating,korotkov2011flying}, acoustic waves~\cite{dumur2021quantum} or electrons\cite{yamamoto2012electrical}, among which photonic flying qubits are most widely used in practice. Depending on the wavelength, photonic flying qubits may work in the optical regime when being used for long-distance communication~\cite{mirhosseini2020superconducting}. They can also be generated and manipulated in the microwave regime for interconnection with solid-state quantum information processing devices~\cite{ilves2020demand}. In this paper, we are mainly concerned with microwave flying qubits, for which superconducting quantum systems are excellent candidates for emitters and receivers for their frequency tunability, long coherence time and field confinement to one-dimensional transmission lines without additional spatial pattern matching~\cite{gu2017microwave}. 

From the viewpoint of quantum input-output theory, flying qubits can be treated as the quantum input of a receiver or quantum output of an emitter. Existing studies have shown that flying qubits sent from an {emitter} must be well shaped to match to the receiver, so that they can be completely absorbed~\cite{2013Microwave,forn2017demand,li2022control}. This gives rise to the flying-qubit shaping problem to be studied in this paper. 

In most flying-qubit generation systems, the emitter is coupled to the waveguide through an intermediate cavity for the enhancement of emitter-photon interaction. The cavity also facilitates the shaping control by varying the cavity's instantaneous photon emission rate~\cite{pierre2014storage,yin2013catch} or the effective emitter-cavity coupling induced by a coherent driving field on the cavity~\cite{ilves2020demand,2013Microwave}. However, the cavities inherently limit the transmission bandwidth of flying qubits and require strict frequency alignment. Moreover, the cavity may also cause random release of unwanted photons. 

Regarding these issues, the cavity-free system using tunable emitters and couplers~\cite{forn2017demand,hoi2011demonstration,hoi2012generation,zhong2019violating,zhong2021deterministic} offers a more efficient scheme~\cite{peng2016tuneable,zhou2020tunable}. Theoretical studies indicate that flying qubits can be tailored into arbitrary shape using an ideal two-level emitter and an ideal tunable coupler~\cite{Yao2005,Nurdin2016,Li2022prb}, which has been experimentally demonstrated in various experiments in superconducting systems~{\color{red}\cite{zhong2021deterministic,qiu2023deterministic,kannan2023demand}}. 

{\color{red} However, realistic emitters and couplers are always non-ideal. In superconducting systems, the transmon qubit is frequently used due to its long lifetime; however, it cannot be regarded as a true two-level system because of its weak anharmonicity. This limitation can result in significant control errors caused by level leakage~\cite{zhou2020tunable}. Regarding the emitter, a very strong coupling beyond its tunable range may be required when the desired photon shape has a sharply rising part. For example, exponentially rising flying qubits are often preferred in practice because they can be fully captured by the receiver without the need for a tunable coupler~\cite{Nurdin2016}, but such flying qubits necessitates infinitely strong coupling that is unattainable with any available couplers~\cite{chen2014qubit,forn2017demand,yan2018tunable,kurpiers2018deterministic,sete2021floating}.}

As a result, additional controls must be implemented when the coupler fails to provide sufficient tunability. It seems that the only option available is the coherent driving field, which is typically employed to prepare the emitter's state but is not preferred for shaping due to the undesirable multi-photon emissions it produces. Nevertheless, we will demonstrate in this paper that coherent control can be designed to improve shaping performance by utilizing quantum optimal control theory, which has been proven highly effective in the manipulation of stationary qubits.~\cite{khaneja2005optimal,reich2012monotonically,boscain2021introduction}. To the best of our knowledge, there are no such investigations in the literature except in linear quantum systems~\cite{zhang2012response}.

The subsequent sections will be organized as follows. Section~\ref{Sec:Model} introduces the mathematical model of the flying-qubit control system implemented by a transmon-qubit emitter, based on which the objective functionals for three associated flying-qubit control tasks are introduced in Sec.~\ref{Sec:OCT}. In Sec.~\ref{Sec:result}, the gradient-descent algorithm subject to the proposed objective functionals are numerically tested. Finally, in Sec.\ref{Sec:Conclusion}, conclusions are {drawn}. 

\section{The modeling of flying-qubit generation systems}\label{Sec:Model}

To illustrate how optimal control theory can effectively address various non-idealities in the flying-qubit shaping system, we assume that the emitter is a superconducting transmon qubit that cannot be treated as a perfect two-level system. The design methodology can be readily adapted to other types of emitters as well.

\begin{figure} \centering
	\includegraphics[width=1\columnwidth]{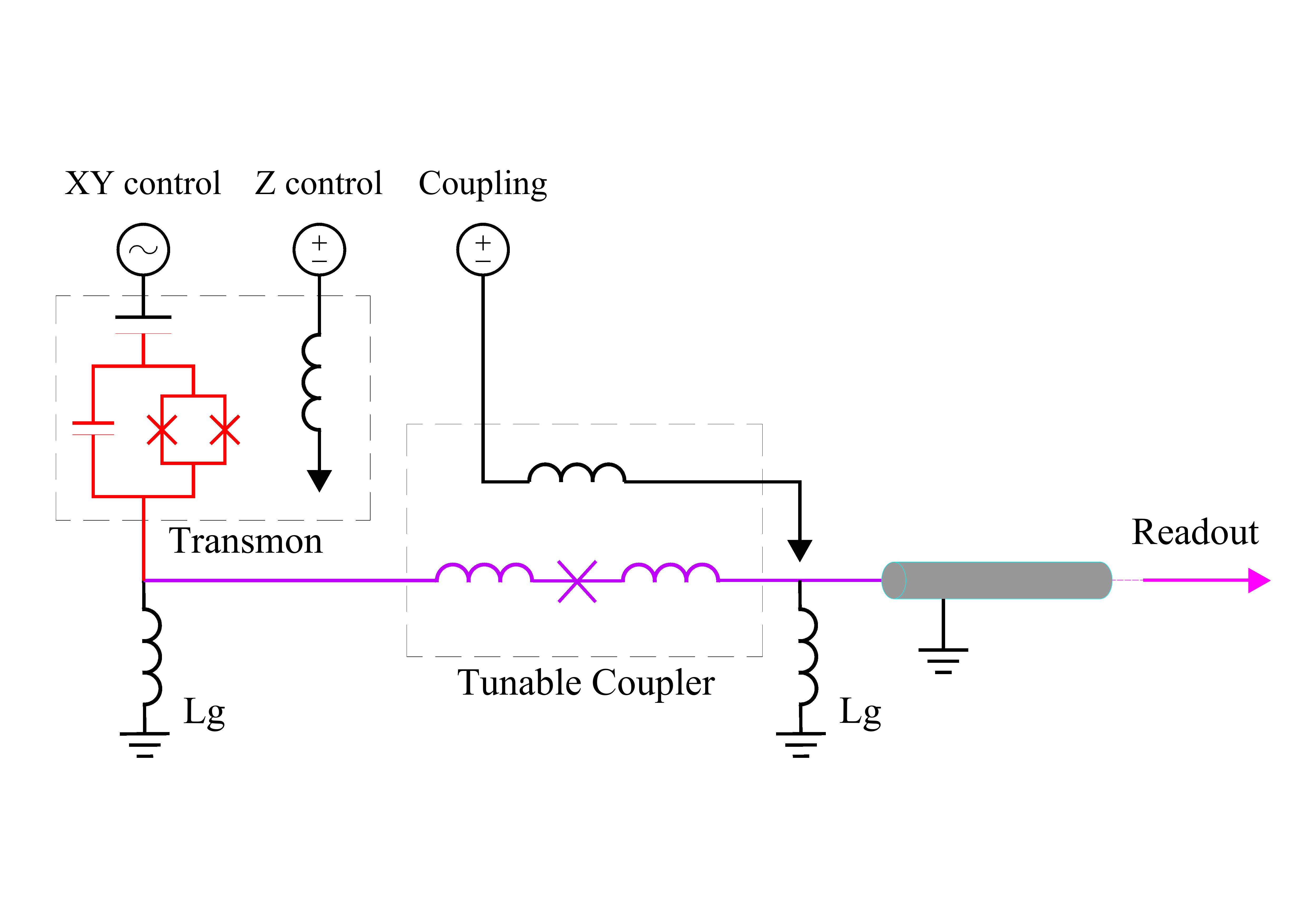}
	\caption{ Schematics of a microwave flying-qubit emitter implemented by a transmon qubit. The emitter is inductively coupled to a transmission-line waveguide through a tunable gmon coupler. The transmon-qubit is frequency tunable by the $Z$ control and its state can be coherently manipulated by the microwave $XY$ driving field.}
	\label{fig1}
\end{figure}

As is shown in Fig.~\ref{fig1}, the transmon-qubit emitter is inductively connected to a unidirectional waveguide (transmission line) via a gmon coupler that the coupling strength can be altered in real time~\cite{chen2014qubit,zhong2021deterministic}. The frequency of the transmon-qubit emitter is tuned by the Z-control line and its state transition can be manipulated by the XY microwave driving field. 

Let $|{\rm vac}\rangle$ be the vacuum state of the photon field in the waveguide, and $\hat{b}(t)$ be its temporal annihilation operator that satisfies the singular commutation relation $[\hat{b}(t),\hat{b}^{\dag}(t')]=\delta(t-t')$~\cite{loudon2000quantum,zhang2012response}. In this paper, we are concerned with the generation of shaped flying qubits, and hence the waveguide can be assumed to be empty when $t<0$. A pulsed single-photon state is defined as follows:
\begin{equation} \label{eq:1-photon}
|1_\xi\rangle = \int_{0}^\infty \xi(t) \hat{b}^{\dag}(t) {\rm d} t|{\rm vac}\rangle,
\end{equation}
in which the complex-number valued function $\xi(t)$
represents the temporal shape of the single-photon pulse. The normalization property of $\ket{1_{\xi}}$ requires that the shape function $\xi(t)$ satisfies:
\begin{equation}
	\int_0^\infty \left|\xi(t)\right|^2{\rm d}t=1.
\end{equation}

In this paper, we assume that the emitter is coupled to the waveguide with $100\%$ efficiency, meaning that the emitted photons all enter the waveguide. A straightforward two-stage protocol for generating flying qubits is to first drive the transmon qubit to some desired superposition state, say $c_0\ket{0}+c_1\ket{1}$, and then transfer it to the photon field state $c_0\ket{\rm vac}+c_1\ket{1_{\xi}}$ via spontaneous emission. The single-photon shape $\xi(t)$ is determined by the real-time tuning of the coupler. Here, the vacuum state $\ket{\rm vac}$ and the single-photon state $\ket{1_\xi}$ represent the logical states $``0"$ and $``1"$, respectively, of the flying qubit. Thus, to fully characterize the flying-qubit state, it is essential not only to identify the logical state but also to incorporate the shape function. 

Under general circumstances, the photon field may also involve multi-photon components due to imperfect state preparation or the presence of a coherent driving field. The corresponding flying-qubit state can be expressed in the following form:
\begin{equation}\label{eq:superposition}
 |{\Xi}\rangle=\xi^{(0)}|{\rm vac}\rangle+
 \int_{0}^\infty \xi^{(1)}(t) b^{\dag}(t) {\rm d}t|{\rm vac}\rangle+|{\Xi^{'}}\rangle,
\end{equation}
where $\xi^{(0)}$ is the probability amplitude of the vacuum state, $\xi^{(1)}(t)$ is the shape of the single-photon component, and $|{\Xi^{'}}\rangle$ represents the rest multi-photon components. Because the full state $\ket{\Xi}$ is always normalized, we have
\begin{equation}
	\left|\xi^{(0)}\right|^2+\int_0^\infty \left|\xi^{(1)}(t)\right|^2{\rm d}t \leq 1.
\end{equation}

In the literature, the few-photon emission process induced by time-variant controls (including coherent driving and incoherent tunable coupling) has been studied using scattering theory~\cite{Fischer2018,Trivedi2018} or quantum input-output theory based on quantum stochastic differential equations (QSDE)~\cite{li2022control}. The developed mathematical models enable the calculation of the flying-qubit state as well as the errors induced by multi-photon emission. Here, we leverage the model developed in our previous work~\cite{li2022control}, which is governed the following non-unitary Schrodinger equation:
\begin{equation}\label{eq:definition of propagator}
\dot{ V}(t) = -i H_{\rm eff}(t) V(t),\quad V(0) = \mathbb{I}
\end{equation}
where $V(t)$ is the non-unitary evolution operator and 
\begin{equation}\label{eq:Heff}
	H_{\rm eff}(t) = \frac{\eta}{2}\hat{a}^{\dagger 2} \hat{a}^2+ u(t) \hat{a}^\dagger+u^*(t)\hat{a}-\frac{i\gamma(t) \hat{a}^\dagger \hat{a}}{2}	
\end{equation}
is the effective Hamiltonian. Here, the transmon qubit is treated as an anharmonic oscillator, where $\hat{a}$ is the annihilation operator and $\eta$ is the anharmonicity. The time-variant function $\gamma(t)$ represents the tunable coupling strength produced by the coupler to the waveguide. The coherent driving field $u(t)=u_x(t)+iu_y(t)$ includes the in-phase and phase-quadrature components $u_x(t)$ and $u_y(t)$. 

The flying-qubit state is determined by vacuum and single-photon components, $\xi^{(0)}$ and $\xi^{(1)}(t)$, in Eq.~\eqref{eq:superposition}. Suppose that the emitter is initially prepared at state $|\psi_{0}\rangle$ and the coherent control $u(t)$, if present, persists for at most a finite duration time. In this case, the emitter eventually decays to its ground state $|0\rangle$ after the flying qubit is released. The flying-qubit state can be calculated as follows~\cite{li2022control}:
\begin{eqnarray}
\xi^{(0)}&=&\langle 0\vert G(\infty,0)|\psi_{0}\rangle, \label{eq:wavepacket0} \\
\xi^{(1)}(t)&=&\sqrt{\gamma(t)}\langle 0\vert G(\infty,t)\hat{a}G(t,0)|\psi_{0}\rangle,\label{eq:wavepacket}
\end{eqnarray}
where
\begin{equation}
	G(t,t')=V(t)V^{-1}(t')	
\end{equation}
is the propagator of \eqref{eq:definition of propagator} from time $t'$ to $t$. The intervention of $\hat{a}$ in Eq.~(\ref{eq:wavepacket}) indicates the occurrence of a quantum jump at time $t$ that dumps the qubit to its lower-level state, and releases a photon into the waveguide. 

As a special case, the single-photon component has the following analytic form~\cite{li2022control}
\begin{equation}\label{eq:xit}
	\xi(t)=\sqrt{\gamma(t)}\exp\left[{-\frac{1}{2}\int_0^t \gamma(\tau){\rm d}\tau}\right]
\end{equation}
when $u(t)=0$ and the initial state is $\ket{\psi_0}=\ket{1}$, which is fully dependent on the tunable coupling $\gamma(t)$. In general cases when $u(t)\neq 0$, the single-photon component has to be numerically calculated.

Note that the multi-photon components $|{\Xi^{'}}\rangle$ in Eq.~(\ref{eq:superposition}), which contribute to control errors, need not to be {explicitly} calculated in the optimal control design. This is because the optimization aims at maximizing the proportion of vacuum and single-photon components, which naturally leads to the minimization of the unwanted multi-photon components, given that the full state $\ket{\Xi}$ is always normalized.

\section{Optimal control theory for the shaping of flying qubits}\label{Sec:OCT}

The above model indicates that the logical state and the photon shape of the flying qubits can be altered by the tunable coupling $\gamma(t)$ or the driving field $u(t)=u_x(t)+iu_y(t)$, which play the role of incoherent \textcolor{blue}{or} coherent controls in the flying-qubit shaping. As will be seen later, they can be jointly applied and optimized to improve the shaping performance with non-ideal emitters and tunable couplers.

In this section, we will discuss how to formulate the optimal control problems for the generation of shaped flying qubits. 
 
\subsection{Preparation of the standing qubit state}
Consider the state preparation of the emitter's state that is to be transferred to the flying qubit. Assume that the emitter is initialized at $\ket{\psi_0}=\ket{0}$ and the target state is $\ket{\psi_{P}}$, and the coupler is non-ideal in that it cannot be completely turned off. 

Let $\gamma(t)=\gamma_0\neq0$ be the residual coupling strength and the coherent control function $u(t)$ is to be optimized to minimize the state-preparation error. We present two alternative objective functionals. 

The first objective functional is defined as the state-preparation error of the transmon qubit
\begin{equation}\label{ME}
J_1^{\rm ME}[u(t)]= 1-\langle \psi_{P}|\rho(T_0)|\psi_{P}\rangle 
\end{equation}
where $T_0$ is the duration time of the control pulse, and the density matrix $\rho(t)$ obeys the following master equation (ME):
\begin{eqnarray}
  \dot{\rho}(t)&=-i\left[\frac{\eta}{2}\hat{a}^{\dagger 2} \hat{a}^2+ u(t) \hat{a}^\dagger+u^*(t)\hat{a},\rho(t)\right] \nonumber+\\  
  & \gamma_0 \left[\hat{a}\rho(t) \hat{a}^\dag-\frac{1}{2}\rho(t) \hat{a}^\dag \hat{a} -\frac{1}{2}\hat{a}^\dag \hat{a} \rho(t)\right]. \label{eq:Master Equation}
\end{eqnarray}
The master equation can be derived by averaging over the flying-qubit state with the waveguide being taken as a Markovian environment~\cite{lindblad1976generators,breuer2002theory}. The goal is to maximize the overlap between the emitter state and the target state, thereby minimizing the leakage to higher non-computational states.

The second objective functional is defined as the control error resulting from photon leakage, because the accompanied decay of excited states alters the population of the computational states. The photon leakage is quantified by the probability of detecting photons in the waveguide before the emitter is steered to the target state $\ket{\psi_P}$:
\begin{equation}\label{QSDE}
J_1^{\rm QSDE}[u(t)]=1-|\langle \psi_{P}|G(T_0,0)|0\rangle|^2,
\end{equation}
where $G(T_0,0)=V(T_0)$ can be obtained from Eq.~\eqref{eq:definition of propagator}. The superscript ``QSDE" comes from quantum stochastic differential equation (QSDE)~\cite{li2022control} that was used to derive Eq.~(\ref{eq:definition of propagator}). 

\subsection{Direct Generation of shaped flying qubits}
Suppose that the emitter has been prepared at the target state. With an ideal coupler, the follow-up release of shaped flying qubits can be done with the following tuning scheme: 
\begin{equation}\label{eq:gammat_solution}
\gamma(t)=\frac{|\xi_{0}(t)|^2}{\int_t^\infty |\xi_0(\tau)|^2{\rm d}\tau},	
\end{equation}
where $\xi_0(t)$ is the target shape of the single-photon component. This formula can be derived from Eq.~\eqref{eq:xit}.

Now we consider the scenario that the coupler cannot be completely turned off or arbitrarily strengthened. In this context, we can introduce $u(t)$ as an auxiliary control alongside $\gamma(t)$, and unify the two-stage flying-qubit generation protocol into a single optimal control task. The corresponding objective functional is defined as the distance between the generated and target flying-qubit states:
\begin{equation}\label{flying qubits}
J_{2}[u(t),\gamma(t)]=|\xi^{(0)}-{c_0}|^2 +\int_{0}^{\infty}|\xi^{(1)}(t)-{c_1}\xi_0(t)|^2dt,
\end{equation}
in which the vacuum and single-photon components of the emitted photon field are 
\begin{eqnarray}
	\xi^{(0)}&=&\langle 0\vert G(\infty,0)\vert 0\rangle, \label{eq:wavepacket0a} \\
	\xi^{(1)}(t)&=&\sqrt{\gamma(t)}\langle 0\vert G(\infty,t)\hat{a}G(t,0)\vert 0\rangle.\label{eq:wavepacketa}
\end{eqnarray}
Note that we adopt the Euclidean distance here for optimization. In practice, one can alternatively use other distance measures.

\subsection{State transfer from the standing qubit to a shaped flying qubit}
Consider the control scenario that the emitter has been prepared to some unknown state $\ket{\psi}=c_0 \ket{0}+c_1\ket{1}$, and we wish to design a control protocol that transfers the emitter to the flying-qubit state $c_0\ket{\rm vac}+c_1\ket{1_{\xi_0}}$. The control protocol should be independent of speficific values of $c_0$ and $c_1$. 

The state transfer actually forms a SWAP gate operation between the emitter and the flying qubit, which can be decomposed into two separate control tasks: (1) the flying qubit should be in the vacuum state when the emitter is initially {prepared} in state $\ket{0}$; and (2) the flying qubit should be in the single-photon state $\ket{1_{\xi_0}}$ when the emitter is initially prepared in state $\ket{1}$. The sum of the corresponding distances between the generated and target flying-qubit states form the following objective functional
\begin{equation}
J_{3}[u(t),\gamma(t)]=
\left|1-\xi^{(0)}\right|^2 
+\int_{0}^{\infty}|\xi^{(1)}(t)-\xi_0(t)|^2dt,
\label{state_transfer}
\end{equation}
where
\begin{eqnarray}
	\xi^{(0)}&=&\langle 0\vert G(\infty,0)\vert 0\rangle, \label{eq:wavepacket0B} \\
	\xi^{(1)}(t)&=&\sqrt{\gamma(t)}\langle 0\vert G(\infty,t)\hat{a}G(t,0)\vert 1\rangle.\label{eq:wavepacketB}
\end{eqnarray}
Note that the calculation of the single-photon component \eqref{eq:wavepacket0B} is different from that in \eqref{eq:wavepacket0a}, in that it is conditioned on the emitter's initial state $\ket{1}$. 

\section{Simulation Results}\label{Sec:result}
In this section, we perform numerical simulations to test the effectiveness of optimal control in flying-qubit shaping problems. We adopt the L-BFGS quasi-Newton algorithm~\cite{dennis1977quasi} for the optimization subject to the objective functionals proposed in Sec.~\ref{Sec:OCT}, and the calculation of the corresponding gradient vectors can be found in Appendix~\ref{appendix}. 

{\color{red}Considering practical constraints, we impose that the control pulse $u(t)$ is zero at both the initial and final time instances (i.e., $u(0)=u(T)=0$). To ensure smoothness during the optimization, we apply a low-pass filter to $u(t)$. Additionally, the control $u(t)$ is constrained with a bound $B=2\pi\times 80$MHz. A detailed discussion on how these constraints are managed in the optimization algorithm can be found in Appendix~\ref{appendix}.}

In the simulation, the anharmonicity of the transmon-qubit emitter is set to $\eta=-2\pi\times 200$~MHz, and we retain the first five levels in the model \eqref{eq:definition of propagator} to assess the effect of level leakage. The target shape of flying qubits considered in the simulations are among the following three types
\begin{enumerate}
	\item Exponentially decaying: $\xi_0(t)=\sqrt{\alpha}e^{-\alpha t/2}$;
	\item Exponentially rising: $\xi_0(t)=\sqrt{\alpha}e^{\alpha (t-T)/2}$;
	\item Symmetrical: $\xi_0(t)=\sqrt{\alpha}\rm{sech}(\alpha \emph{t}/2)$, 
\end{enumerate}
where $\alpha$ {determines} the width of the shape function. All the shaped functions are defined on a sufficiently large time interval $[0,T]$.

\subsection{Flying-qubit control under fixed coupling and tunable coherent control}
We first investigate the capability of coherent control $u(t)$ on the shaping of flying qubits with a fixed coupler. 
\subsubsection{Preparation of standing transmon-qubit state}
Assume that the coupler's residue coupling strength is $\gamma_0= 2\pi\times 0.5$MHz. We optimize the coherent control $u(t)$ subject to the two objective functionals \eqref{ME} and \eqref{QSDE}, in which the target emitter state is specified as $\ket{\psi_P}=\ket{1}$ (i.e., $c_0=0$ and $c_1=1$) with a pulse duration $T_0=10$ns.

\begin{figure} \centering
	\includegraphics[width=0.95\columnwidth]{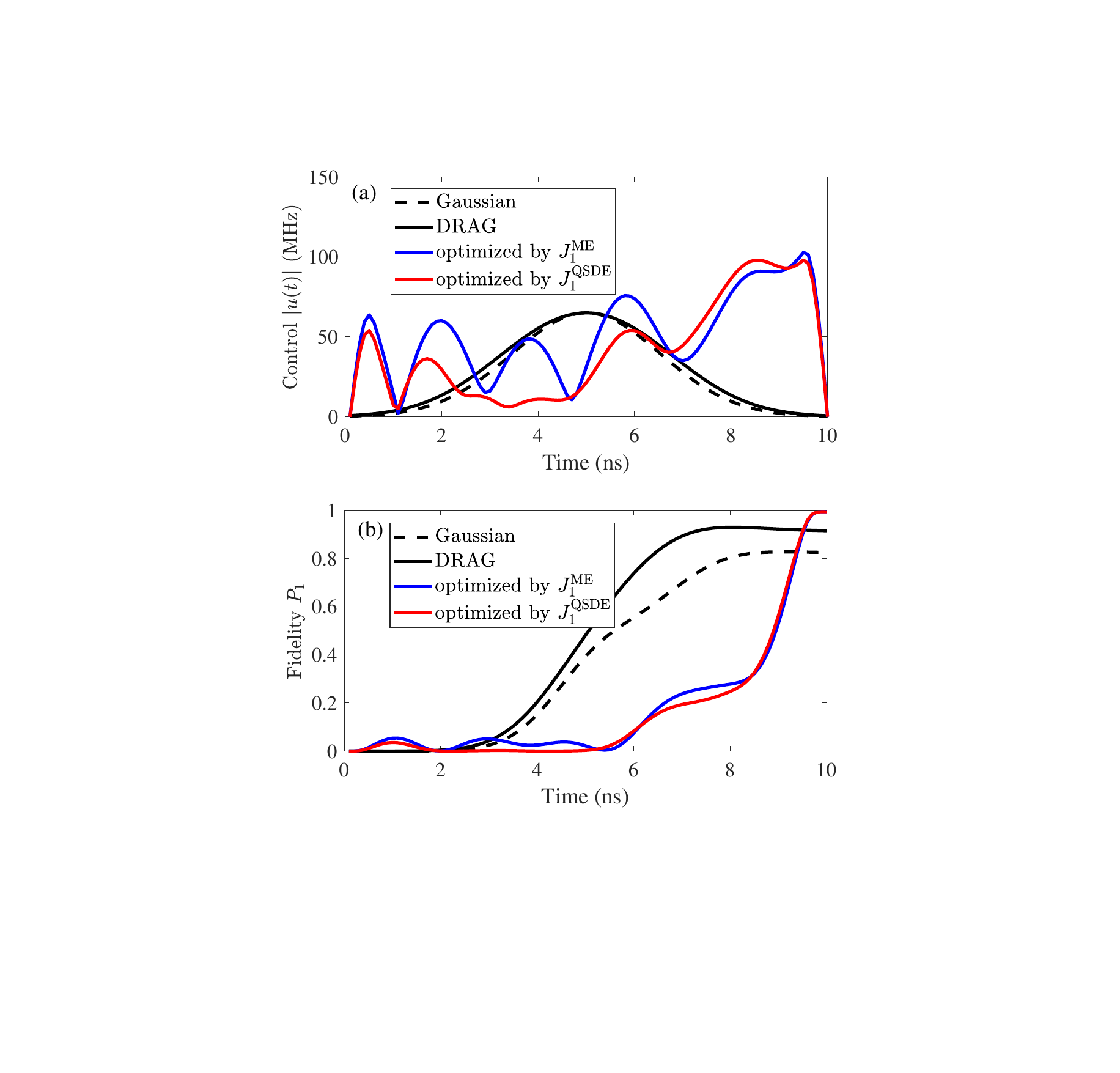}
	\caption{(a) The amplitude of the Gaussian $\pi$-pulse, the DRAG pulse and the optimized control pulses subject to objective functionals $J_1^{\rm ME}$and$J_1^{\rm QSDE}$; (b) {the fidelity of the transmon-qubit being in the target state $\ket{1}$ during the control process.}}
	\label{fig2}
\end{figure}

For comparison, we also simulated the standard Gaussian $\pi$-pulse that is commonly used for state flipping in ideal two-level systems, and its DRAG correction~\cite{Motzoi2009} for level-leakage suppression. Figure~\ref{fig2}(a) displays the amplitudes of these control fields. {\color{red}The control optimized with $J_1^{\rm ME}$ contains oscillating components that manage the unwanted state transitions associated with the level leakage. In contrast, the control optimized with $J_1^{\rm QSDE}$ exhibits a more concentrated pulse energy towards the end of the pulse. This suggests that the optimization strategy seeks to keep the emitter unexcited for as long as possible to reduced the photon leakage.}

Figure~\ref{fig2}(b) illustrates the population variation of the target state $\ket{1}$ throughout the control process. At the final time $t=T_0$, the population reaches $\ket{1}$ with $99.40\%$ and $99.41\%$ fidelities, respectively. The DRAG pulse achieves a lower fidelity $91.56\%$, while the Gaussian pulse yields the poorest performance with only $82.52\%$ fidelity. 

\begin{figure} \centering
	\includegraphics[width=1\columnwidth]{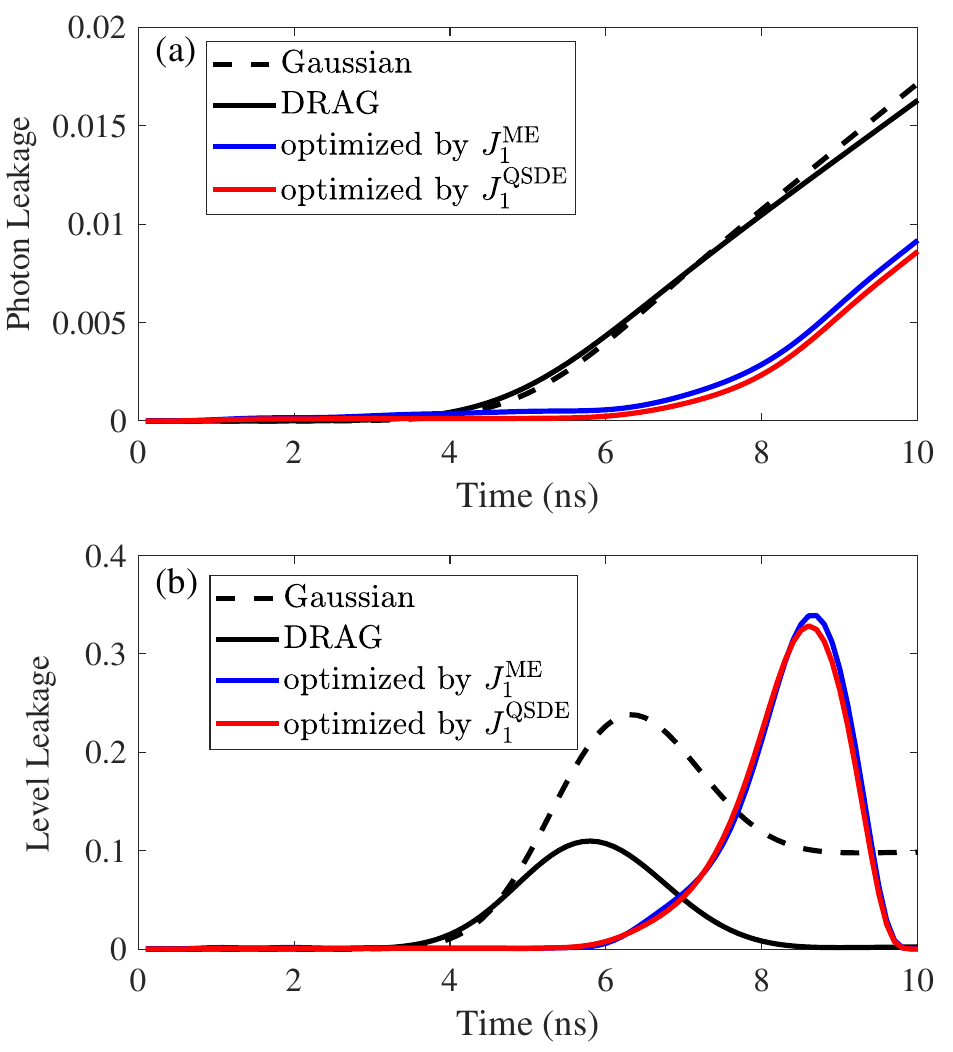}
	\caption{(a) The photon leakage $1-|\xi^{(0)}|^2$ and (b) the level leakage (total population $P_2+P_3+P_4$ of non-computational states $\ket{2}$, $\ket{3}$ and $\ket{4}$) during the state preparation. }
	\label{fig3}
\end{figure}

To identify the sources of errors, we analyze in Fig.~\ref{fig3} the corresponding photon leakage (i.e., the total population $1-|\xi^{(0)}|^2$ of non-vacuum field states) and the level leakage (i.e., the total population of non-computational states $\ket{2}$, $\ket{3}$ and $\ket{4}$). In Fig.~\ref{fig3}(a), the photon leakage monotonically increases with time in all cases owing to the Markovian dynamics. The increase is significantly slower under the optimal controls, indicating that the photon leakage is lower compared to the {Gaussian} and DRAG controls. Figure~\ref{fig3}(b) demonstrates that the level leakage can be nearly completely suppressed by using the DRAG pulse and the optimized pulses, whereas the performance of Gaussian pulse is considerably inferior. 

The error analysis provides a clear {explanation} for the performance differences observed in Fig.~\ref{fig2}(b). The two optimal controls achieve high fidelities because they effectively mitigate both the photon leakage and the level leakage. While the DRAG pulse successfully reduces errors caused by level leakage, it does not efficiently suppress the photon leakage error that results in population loss of $\ket{1}$ that decays to $\ket{0}$. 

The simulation results are also interesting in that, although the objective functionals $J_1^{\rm ME}$ is designed for level-leakage suppression, the resulting optimal control also effectively reduces the photon leakage. Similarly, the optimal control obtained from $J_1^{\rm QSDE}$ can suppress both types of leakage errors as well. {\color{red}We conjecture that the two objective functionals may be equivalent in the context of state preparation in open systems, but this remains to be confirmed in our future studies.} 

\subsubsection{Direct generation of shaped flying qubits}\label{sec:generation}
Consider the direct generation of flying qubits by optimizing $u(t)$ subject to $J_2$ defined in Eq.~\eqref{flying qubits}, where the coupling strength $\gamma(t)$ is fixed at $\gamma_c=2\pi\times5$MHz. The exponentially decay, exponentially rising and symmetric target single-photon shapes are all tested with the width parameter $\alpha=2\pi\times 6$MHz. The control duration time is chosen as { $T=600\ {\rm ns}$, which is much longer than the decay time $\gamma_c^{-1}$}. 

\begin{figure} \centering
	\includegraphics[width=   0.95\columnwidth]{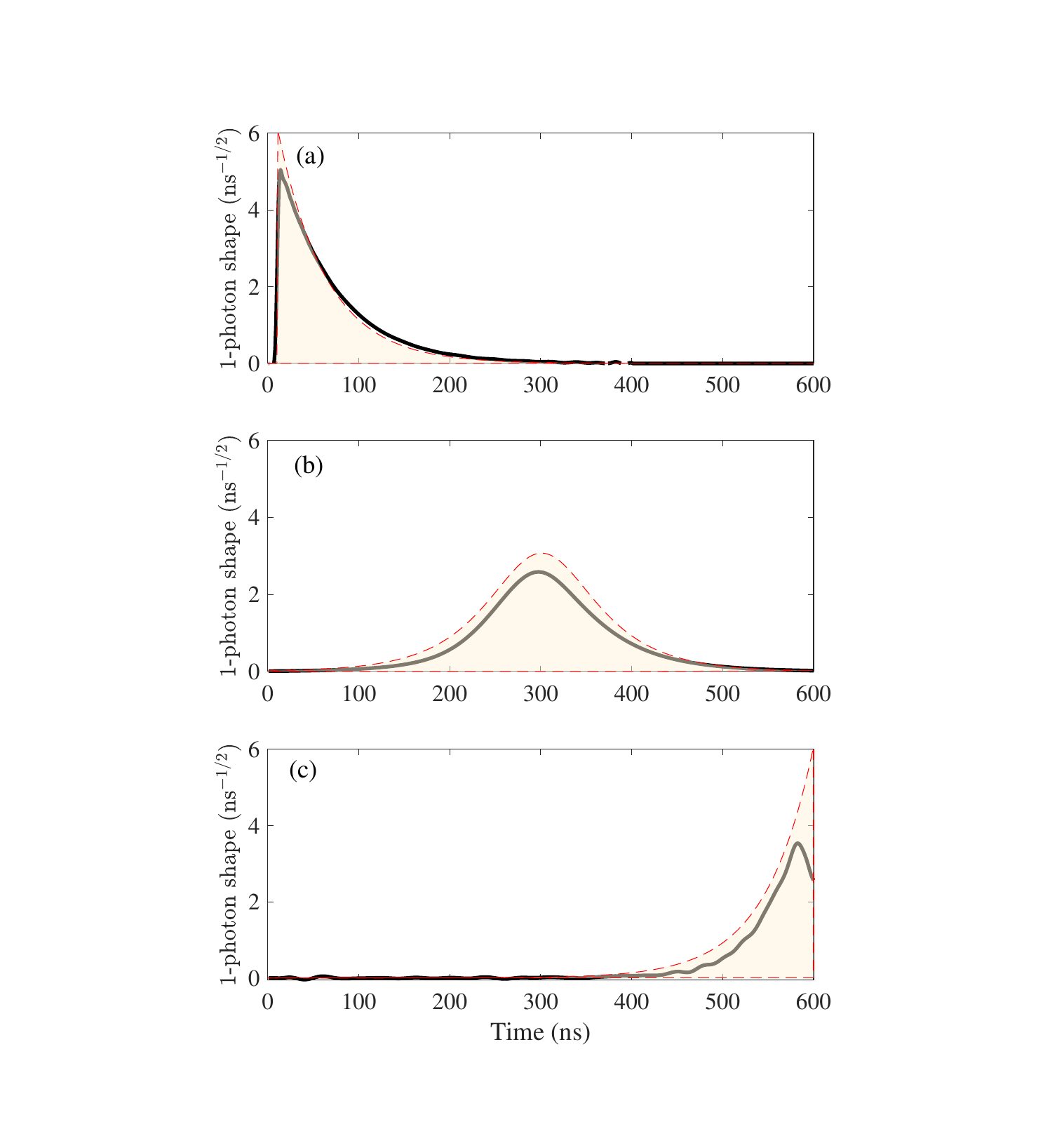}
	\caption{ The optimized shapes of single-photon components under coherent driving control and fixed coupling strength $\gamma_c=2\pi\times 6$MHz, where the target single-photon shapes are exponentially decaying, symmetric and exponentially rising, respectively, with $\alpha=2\pi\times 6$MHz.}
	\label{fig4}
\end{figure}

\begin{figure}
	\includegraphics[width=1\columnwidth]{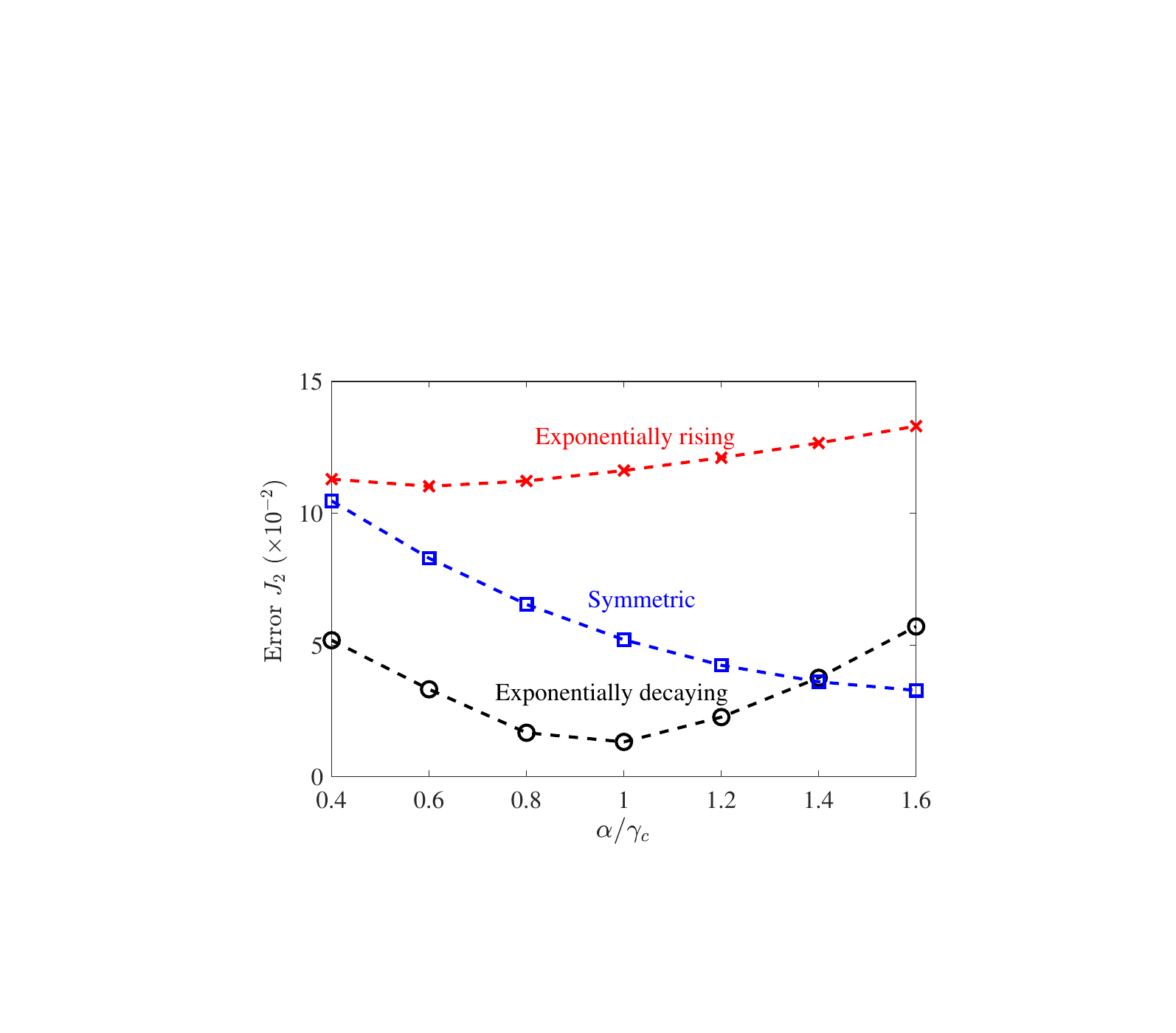}
	\caption{The best performance achieved by coherent control for different types of single-photon shapes under different values of $\alpha$ ranging from $0.4\gamma_c$ to $1.6\gamma_c$ }
	\label{fig5}
\end{figure}

In the simulations, the control pulse $u(t)$ is initiated $10$ns prior to $\xi_0(t)$ when dealing with exponentially decaying flying qubits, because the emitter must be {promptly} excited to $\ket{1}$ before the photon emission occurs. Note that this early initiation is not required for the other two shapes whose beginning part rises slowly. Figure~\ref{fig4} displays the optimized single-photon pulse shapes, which all closely match their target waveforms. We further optimize the controls with $\alpha$ {varying} from $0.6\gamma_c$ to $1.4 \gamma_c$, and plot in Fig.~\ref{fig5} the relationship between the best performance (i.e., the minimal value of $J_2$) and the parameter $\alpha$. 

The simulation results indicate that the coherent control $u(t)$ possesses some capacity for single-photon shaping in the absence of a tunable coupler; however, the control performance is far from good, particularly when the target pulse shape includes a rising segment. Among the three types of shapes, generating exponentially-decaying shape flying qubits is the easiest, as the shape is close to those produced by natural spontaneous emission. In contrast, generating flying qubits with exponentially rising shapes proves to be considerably more challenging.

\subsubsection{State transfer from the emitter to a shaped flying qubit}\label{subsection:transfer}
Now we optimize the coherent control $u(t)$ for the state transfer from the emitter to a flying qubit with exponentially decaying, exponentially rising and symmetric shapes. The parameters are set to $\alpha=2\pi\times 6$MHz and $\gamma_c=2\pi\times5$MHz. 

The optimization is subject to $J_3$ defined by Eq.~\eqref{state_transfer}. To assess the state-transfer performance, we plot the profiles of the single-photon components conditioned on the emitter's initial state. Ideally, a single photon $\ket{1_{\xi_0}}$ should be released (i.e., $\xi^{(1)}(t)=\xi_0(t)$) when the emitter is initiated at $\ket{\psi_0}=\ket{1}$.{ Conversely}, when $\ket{\psi_0}=\ket{0}$, no single-photon emission is expected (i.e., $\xi^{(1)}(t)=0$). 

\begin{figure}
	\includegraphics[width=1\columnwidth]{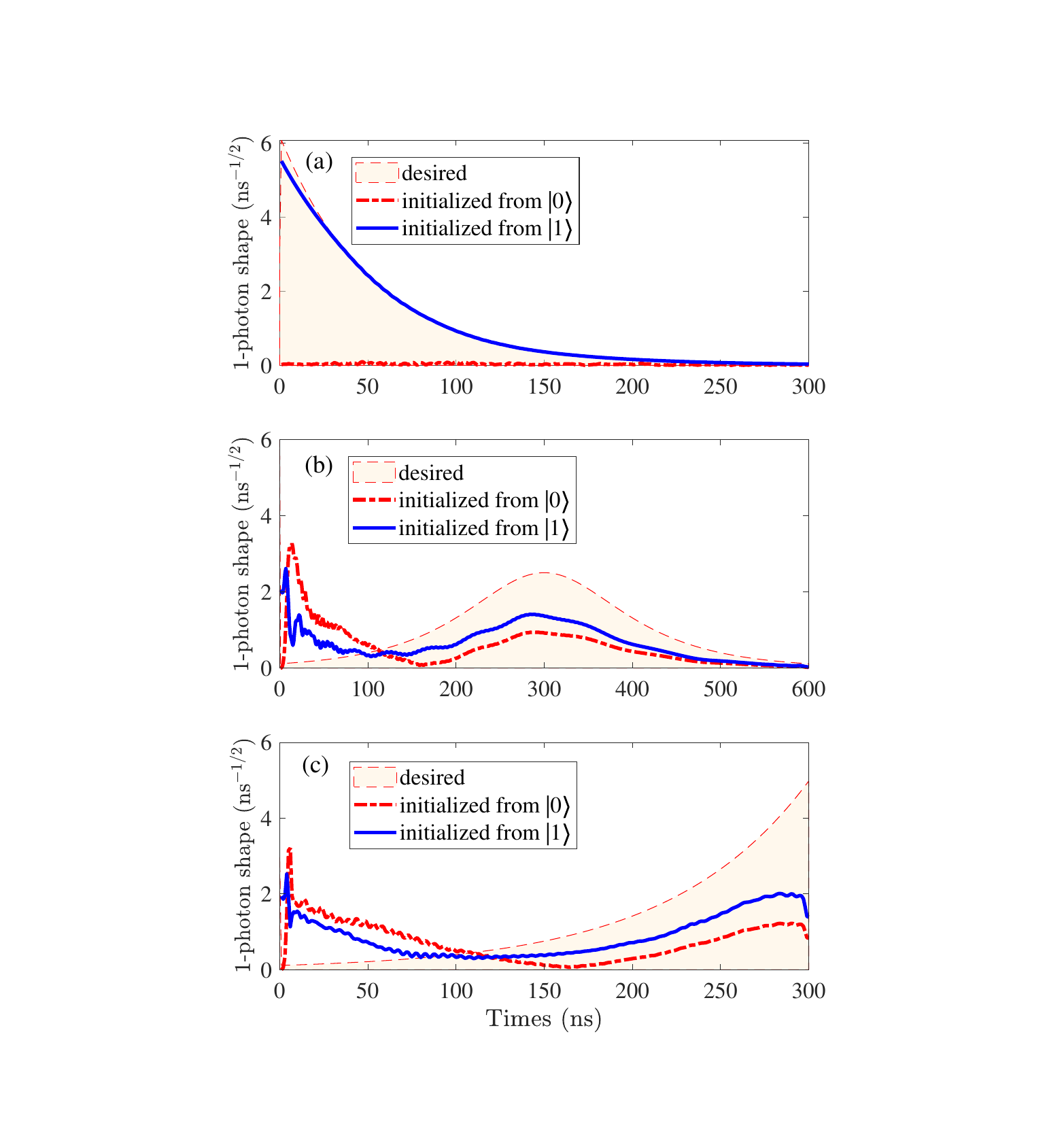}
	\caption{The optimized state transfer from the transmon qubit to flying qubits with (a) exponentially decaying, (b) symmetric, and (c) exponentially rising shape.}
	\label{fig6}
\end{figure}

Figure~\ref{fig6} displays the conditioned single-photon components under optimal controls for the three types of target single-photon shapes. {Similar} to the findings in Sec.~\ref{subsection:transfer}, the state transfer is almost perfect when the target single-photon shape is exponentially decaying. However, the performances is significantly poorer when the target single-photon shape is either symmetric or exponentially rising. The single-photon components conditioned on $\ket{\psi_0}=\ket{0}$ and $\ket{\psi_0}=\ket{1}$ is nearly indistinguishable, falling short of the anticipated conditional single-photon emission. The overall poor performance indicates again that the coherent control $u(t)$ offers limited capacity for single-photon shaping without a tunable coupler.  

\subsection{Flying-qubit control under tunable coupling and tunable coherent control}
The above simulations collectively demonstrate the necessity of a tunable coupler for high-quality flying-qubit shaping. Next, we will demonstrate that, while coherent control is not ideal for flying-qubit shaping, it can serve as a valuable complement to enhance the performance of a tunablility-limited coupler.

We begin by testing the direct generation of flying qubits with an exponentially-rising shape, which is the most {challenging} among the three shapes. In Sec.~\ref{sec:generation}, the {achievable} minimal shaping error is $0.116$ with a fixed-strength coupler ($\gamma_c=2\pi\times5$MHz). On the other hand, according to Eq.~\eqref{eq:gammat_solution}, the shaping can be perfectly done without the coherent control $u(t)$ with an ideal tunable coupler as follows~\cite{li2022control}:
\begin{equation}\label{eq:ideal gamma}
\gamma(t)=\frac{\alpha e^{\alpha(t-T)}}{1-e^{\alpha(t-T)}}.
\end{equation}
Apparently, as can be seen in Fig.~\ref{fig5}(b), the tuning scheme \eqref{eq:ideal gamma} is unrealistic because the desired $\gamma(t)$ approaches infinity when $t\rightarrow T$.

Assume that the coupling strength of the non-ideal tunable coupler is bounded below by $2\pi \times 0.5$MHz and above by $2\pi \times 10$MHz. An approximate tuning scheme [see Fig.~\ref{fig7}(b)] can be derived by simply applying a cut-off to the ideal scheme described in \eqref{eq:ideal gamma}. As illustrated in Fig.~\ref{fig7}(a), the cut-off scheme produces a poorly-shaped single photon, primarily due to the initial photon leakage caused by the inability to fully turning off the coupling. 

Now we maintain the cut-off tuning scheme and introduce the coherent control $u(t)$ as an auxiliary control. As is shown in Fig.~\ref{fig7}(a), optimizing $u(t)$ subject to $J_3$ allows the single-photon component to closely resemble $\xi_0(t)$ except the ending part that requires extremely strong coupling. This optimization process reduces the control error from $0.112$ to $0.038$. If we allow the coupling function $\gamma(t)$ to be jointly optimized alongside $u(t)$, the control error can be further decreased from $0.038$ to $0.028$. 

{\color{red}These improvements indicate that the coherent control is a useful complement to the tunability-limited coupler. As depicted in Fig.~\ref{fig7}(c), the coherent control pulses are active in the regions where the coupler's tunability is restricted (roughly before $530$ns and after $580$ns). Within the time interval where the desired coupling strength is achievable, the coupler primarily governs the system, while the coherent control remains nearly inactive. }

{We also evaluated the optimization of state transfer protocols using the aforementioned schemes involving a tunability-limited coupler, with simulation results presented in Fig.~\ref{fig8}. Compared to the coherent control scheme with a fixed coupler [see Fig.~\ref{fig6}(c)], the cut-off coupling scheme without coherent control results in anticipated zero photon emission when the emitter's initial state is $\ket{\psi_0}=\ket{0}$, but the single-photon component conditioned on $\ket{\psi_0}=\ket{1}$ is still poorly shaped. If we jointly optimize $u(t)$ and $\gamma(t)$, the single-photon shape conditioned on $\ket{\psi_0}=\ket{1}$ is significantly improved. However, the single-photon emission conditioned on $\ket{\psi_0}=\ket{0}$ cannot be fully suppressed. 

To quantitatively compare the control performances, we list in Table~\ref{Exponential} the errors contributed by the conditional state-transfer infidelities
for $\ket{0}\rightarrow \ket{{\rm vac}}$ and $\ket{1}\rightarrow \ket{1_{\xi_0}}$, respectively, 
\begin{eqnarray}
	E_{\vert 0\rangle \rightarrow |{\rm vac}\rangle}&=& 
	\left|1-\xi^{(0)}\right|^2,\\
	E_{\vert 1\rangle \rightarrow \vert 1_{\xi_0} \rangle} &=& \int_{0}^{\infty}\vert \xi^{(1)}(t)-\xi_0(t)\vert ^2{\rm d}t.
\end{eqnarray}
The comparisons demonstrate that the tunable coupler is more effective than coherent driving in shaping control, despite the coupler's limited tunability. Nevertheless, the coherent control plays a crucial complementary role when used in conjunction with the coupler, resulting in a significant reduction of the total error from $0.3445$ to $0.0968$.

\begin{table}[]
	\caption{The error budget under state transfer control schemes: (A) fixed coupling with tunable coherent control, (B) cut-off coupling without coherent control; (C) jointly optimized coupling and coherent control}	
		\label{Exponential}
		\begin{tabular}{c|c|c|c}
			\hline
			Control scheme & $E_{|0\rangle \rightarrow |{\rm vac}\rangle}$  & $E_{|1\rangle \rightarrow |1_{\xi_0} \rangle}$ &  Total error \\ \hline
			(A)  & 0.4534  & 0.3593  & 0.8127  \\ \hline
			(B)  & 0  & 0.3435  & 0.3435  \\ \hline	
			(C)  & 0.0317  & 0.0651  & 0.0968  \\ \hline			
		\end{tabular}
\end{table}

\begin{figure} \centering
	\includegraphics[width=1\columnwidth]{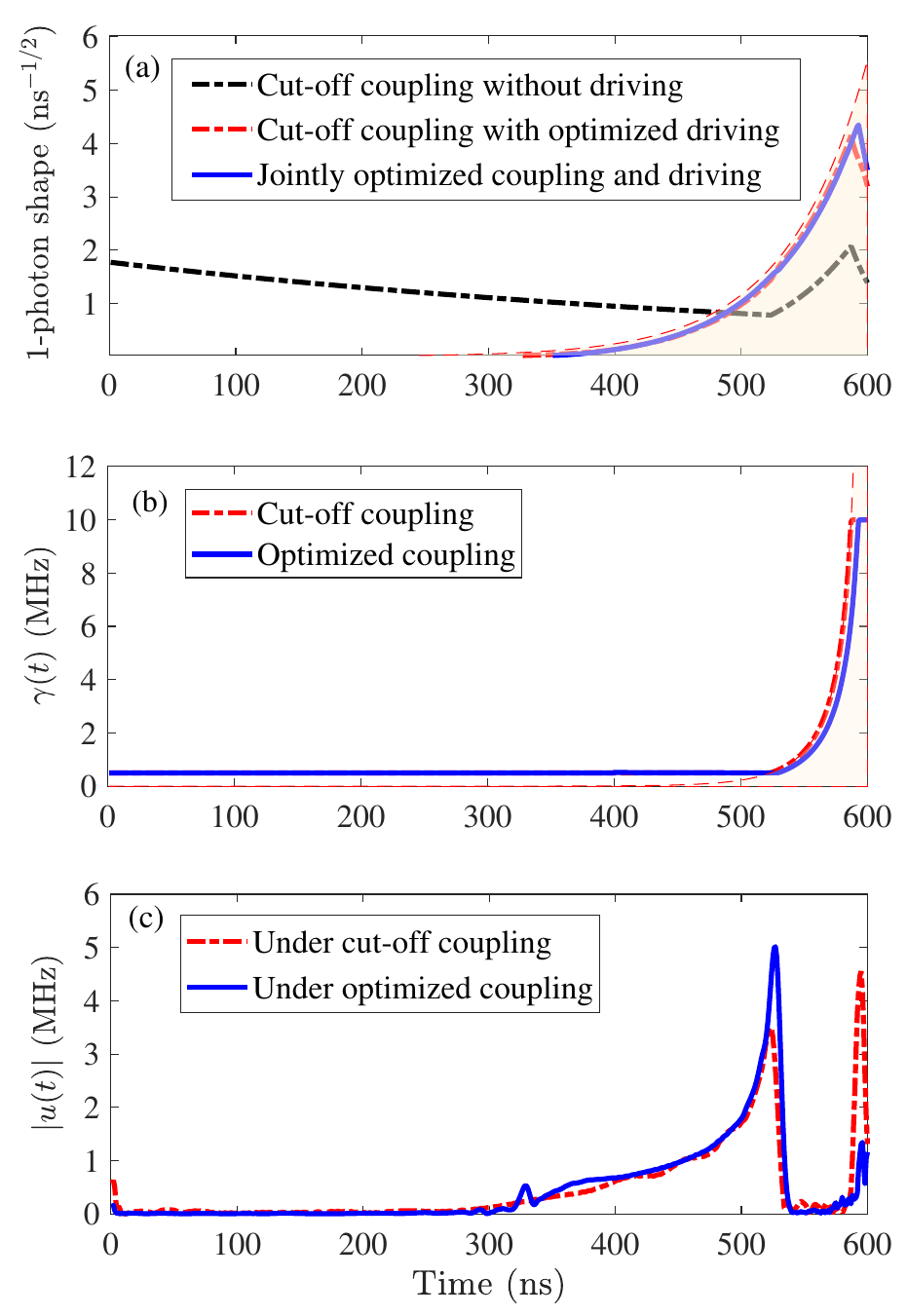}
	\caption{The fitting result of exponentially-rising
		shape flying qubits with the rate $\alpha=2\pi\times5$MHz: (a) single-photon pulse shapes; (b) the corresponding tunable coupling function; (c) the amplitude of the control pulses.}
	\label{fig7}
\end{figure}

\begin{figure} \centering
	\includegraphics[width=1\columnwidth]{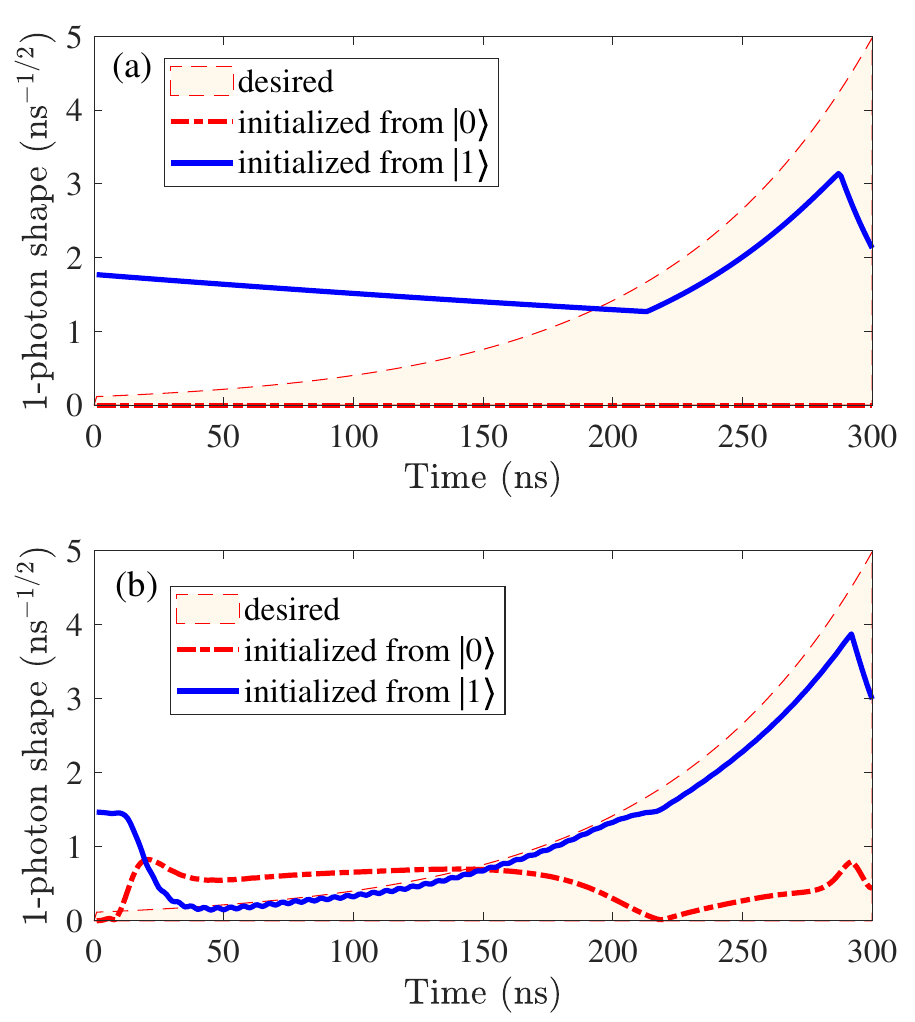}
	\caption{The single-photon emission during the state transfer from the transmon to the flying qubit with exponentially rising shape: (a) under cut-off coupling without coherent control; (b) under optimized coupling and optimized coherent control.}
	\label{fig8}
\end{figure}

\section{Conclusion}\label{Sec:Conclusion}In conclusion, we have introduced quantum optimal control theory for the manipulation of flying qubits. The proposed gradient-descent algorithm is designed for systems involving non-ideal emitters and couplers. Simulation results indicate that the optimized coherent controls can effectively reduce both level and photon leakages caused by the imperfections, but their shaping capabilities are  constrained without a tunable coupler. Nonetheless, coherent control can serve as an important complement to enhance performance, particularly in scenarios where coupler tunability is limited.

Throughout this paper, our discussion focuses on a transmon qubit with a single input-output channel to {facilitate} the physical comprehension. The established design framework can be seamlessly adapted to general flying-qubit control systems with more complicated emitters and multiple input-output channels. Additionally, it is compatible with various optimization methods, including Pontryagin's Minimum Principle~\cite{bao2018optimal} and the Krotov method~\cite{tannor1992control}. In the presence of other non-idealities such as parametric uncertainties and pulse distortions in the control lines~\cite{Cao2022}, control fidelities can be further improved through robust quantum control~\cite{Wu2019} or reinforcement learning techniques~\cite{Sivak2022}.

This study paves the way for a range of intriguing problems in flying-qubit control, such as the generation of entangled or correlated flying qubits for quantum information distribution, as well as the capture and conversion of flying qubits. Additionally, the established framework can be utilized to design flying-qubit-mediated remote quantum gates between distant stationary qubits. These topics will be explored in our future studies.

\appendix
\section{The gradient vector formulas}\label{appendix}
The calculation of gradient vectors is the key part of the optimization algorithm. Here, we present the gradient-vector {formulas} for the proposed objective functionals $J_1^{\rm QSDE}$, $J_2$ and $J_3$, which all rely on the evolution equation \eqref{eq:definition of propagator} . The gradient vector of $J_1^{\rm ME}$ can be found in the literature (e.g., using QuTip ~\cite{johansson2012qutip}), and will not be discussed here. 

To numerically evaluate the objective functionals and their gradients, we choose a sufficiently long time interval $[0,T]$ so that the field emitted after $t=T$ is negligible. The time interval is then evenly discretized into $M$ pieces with $\Delta t=T/M$. Denote $t_k=k\Delta t$, $k=0,1,\cdots,M$, and assume that both $u(t)$ and $\gamma(t)$ are piecewise-constant functions over sub-intervals $[t_{k-1},t_{k}]$, $k=1,\cdots,M$. The transition operator from $t_j$ to $t_n$ can be calculated as
\begin{equation}\label{eq:Ginftytn}
G_{n,j}\triangleq G(t_n,t_j)=V_n \cdots V_{j+2}V_{j+1},
\end{equation}
where $V_k= e^{-i H_{\rm eff} (t_k)\Delta t}$.

According to the first-order perturbation
\begin{eqnarray}
\Delta V_j&\approx&\left[-i\Delta u_x(t_j)(\hat{a}^\dag+\hat{a})+\Delta u_y(t_j)(\hat{a}^\dag-\hat{a})\right.\nonumber \\
&&\left.-\frac{\Delta \gamma(t_j)}{2}
\hat{a}^\dag\hat{a}\right]V_j\Delta t, \label{variation}
\end{eqnarray}
we obtain the elementary partial derivatives
\begin{eqnarray}
\frac{\partial V_k}{\partial u_x(t_j)}&\approx& -i\delta_{jk}(\hat{a}^\dag+\hat{a}) V_{k}\Delta t,\label{eq:dV/dc1}\\
\frac{\partial V_k}{\partial u_y(t_j)}&\approx& \delta_{jk}(\hat{a}^\dag-\hat{a}) V_{k}\Delta t,\label{eq:dV/dc2}\\
\frac{\partial V_k}{\partial \gamma(t_j)}&\approx& -\delta_{jk}\frac{\hat{a}^\dag \hat{a}}{2} V_{k}\Delta t,\label{eq:dV/dc3}
\end{eqnarray}
where $\delta_{jk}$ is the Kronecker symbol.

Based on \eqref{eq:dV/dc1}-\eqref{eq:dV/dc3}, it is easy to derive that 
\begin{eqnarray}\label{grd:QSDE}
\frac{\partial J_1^{\rm QSDE}} {\partial u_x(t_j)}&\approx& -i\langle\psi_P|G_{M_0,j}\left(\hat{a}^\dag+\hat{a}\right) G_{j,0}|0\rangle\Delta t, \\
\frac{\partial J_1^{\rm QSDE}} {\partial u_y(t_j)}&\approx& \langle\psi_P|G_{M_0,j}\left(\hat{a}^\dag-\hat{a}\right) G_{j,0}|0\rangle\Delta t,
\end{eqnarray}
where $j=1,\cdots,M_0$ with $M_0= T_0/\Delta t$.

The calculation of gradient vectors for $J_2$ and $J_3$ is more complicated. Take $J_3$ for example, we can first approximate them as the following summation:
 \begin{eqnarray}
 J_{3}&\approx& |\xi^{(0)}-1|^2 \nonumber \\
 &&+\sum_{n=1}^{M}\left|\xi^{(1)}(t_n)-\xi_0(t_n)\right|^2\Delta t,  \label{eq:obj2}
 \end{eqnarray}  
from which the gradient vector can be calculated as follows
\begin{eqnarray}
	\frac{\partial J_{3}}{\partial {\bf c}(t_j)} &\approx& 2\text{Re}\left\{\frac{\partial \xi^{(0)}}{\partial{\bf c}(t_j)}(\xi^{(0)}-1)^*+\sum_{n=1}^M \frac{\partial \xi^{(1)}(t_n)}{\partial {\bf c}(t_j)}\right.\nonumber\\
	&&\left.\left[\xi^{(1)}(t_n)-\xi_0(t_n)\right]^*\Delta t\right\},
\end{eqnarray}
where ${\bf c}(t)\in\{ u_x(t),u_y(t),\gamma(t)\}$. 

Thus, it is sufficient to compute the derivatives of $\xi^{(0)}$ and $\xi^{(1)}(t)$ with respect to the control variables. Apply the elementary derivatives \eqref{eq:dV/dc1}-\eqref{eq:dV/dc3}, it is not hard to obtain 
\begin{eqnarray} \label{eq:xigradientu}
		\frac{\partial \xi^{(0)}}{\partial u_x(t_j)}&\approx& -i\langle 0\vert G_{M,j}\left(\hat{a}^\dag+\hat{a}\right) G_{j,0}|0\rangle\Delta t,\\
		\frac{\partial \xi^{(0)}}{\partial u_y(t_j)}&\approx& \langle 0\vert G_{M,j}\left(\hat{a}^\dag-\hat{a}\right) G_{j,0}\vert 0\rangle\Delta t,\\
		\frac{\partial \xi^{(0)}}{\partial \gamma(t_j)}&\approx& -\frac{1}{2}\langle 0\vert G_{M,j}{\hat{a}^\dag \hat{a}}G_{j,0}\vert 0\rangle\Delta t.\label{eq:d xi0/d gamma} 
\end{eqnarray}
and
\begin{widetext}
	\begin{eqnarray}
	\frac{\partial \xi^{(1)}(t_n)}{\partial u_x(t_j)}&\approx& \left\{\begin{aligned}
		-i\sqrt{\gamma(t_n)} \langle 0\vert G_{M,n}\hat{a}G_{n,j}\left(\hat{a}^\dag+\hat{a}\right) G_{j,0}\vert 1\rangle\Delta t,\quad t_j \leqslant t_n,\\
		-i\sqrt{\gamma(t_n)} \langle 0\vert G_{M,j}\left(\hat{a}^\dag+\hat{a}\right)G_{j,n}\hat{a}G_{n,0}\vert 1\rangle\Delta t,\quad t_j>t_n.
		\end{aligned}\right. \\
				\frac{\partial \xi^{(1)}(t_n)}{\partial u_y(t_j)}&\approx& \left\{\begin{aligned}
		~~~\sqrt{\gamma(t_n)} \langle 0\vert G_{M,n}\hat{a}G_{n,j}\left(\hat{a}^\dag-
		\hat{a}\right) G_{j,0}\vert 1\rangle\Delta t,\quad t_j \leqslant t_n,\\
		~~~\sqrt{\gamma(t_n)} \langle 0|G_{M,j}\left(\hat{a}^\dag-\
		\hat{a}\right) G_{j,n}\hat{a}G_{n,0}|1\rangle\Delta t,\quad t_j>t_n.
		\end{aligned}\right.\\
		\frac{\partial \xi^{(1)}(t_n)}{\partial \gamma(t_j)} &\approx& \left\{\begin{aligned}
			-\frac{\sqrt{\gamma(t_n)}}{2}\langle 0\vert G_{M,n}\hat{a}G_{n,j}\hat{a}^\dag\hat{a}G_{j,0}|\psi_{0}\rangle\Delta t,\quad t_j < t_n,\\
	-\frac{\sqrt{\gamma(t_n)}}{2}\langle 0|G_{M,j}\hat{a}\left[\hat{a}^\dag\hat{a}-\gamma^{-1}(t_j)\right]G_{j,0}\vert 1\rangle\Delta t,\quad t_j = t_n,\\
	-\frac{\sqrt{\gamma(t_n)}}{2}\langle 0|G_{M,j}\hat{a}^\dag\hat{a}G_{j,n}\hat{a}G_{n,0}\vert 1\rangle\Delta t,\quad t_j>t_n.
	\end{aligned}\right.\label{eq:d xi1/d gamma}
	\end{eqnarray}
\end{widetext}

The gradient-vector calculation of $J_3$ defined by \eqref{flying qubits} is very similar to that of $J_2$ described above, and the corresponding formulas will not be provided here.

{\color{red} In the simulations, we incorporate practical constraints on the coherent control variables. Let $\vec{u}_{x,y}=[u_{x,y}(t_0,\cdots,t_M)]$ be the vectors of variables to be optimized. We begin by selecting a smooth envelop function $s(t)$ with $s(t_0)=s(t_M)=0$ and a linear low-pass filter that can be represented by a matrix $F_{\rm LP}$ acting on $\vec{u}_{x,y}$. We choose control vectors in the following form:
\[\vec{u}_{x,y}=S_{\rm en}F_{\rm LP} \vec{v}_{x,y}\]
in $\vec{v}_{x,y}$ is unconstrained and \[S_{\rm en}={\rm diag}\{s(t_0),\cdots,s(t_M)\}.\] By this means, the control amplitudes always start and end at zero values and are smoothed out during the optimization. 

The corresponding optimization problem can be easily transformed into an unconstrained problem with respect to $\vec{v}_{x,y}$. The gradient vector can be computed as follows:
\[\frac{\partial J}{\partial \vec{v}_{x,y}}=(S_{\rm en}F_{\rm LP} )^\top\frac{\partial J}{\partial \vec{u}_{x,y}},\]
where $\frac{\partial J}{\partial \vec{u}_{x,y}}$ can be calculated from the above derivations.

We also impose a bound $B$ on the control amplitudes. Accordingly, the gradient formulas remain unchanged when $|u_{x,y}(t_j)|<B$. When the bound is saturated, we need to set $\frac{\partial J}{\partial {u}_{x,y}(t_j)}=0$.}
\bibliography{FQshaping}

\end{document}